\documentclass[dvips]{article}
\usepackage{icrctc07}

\title{Simultaneous H.E.S.S. and Chandra observations of Sgr A$^{\star}$ during an X-ray flare}
\shorttitle{Observations of Sgr A$^{\star}$ during an X-ray flare}

\authors{Jim Hinton$^{1}$, Matthieu Vivier$^{2}$, 
Rolf B\"uhler$^{3}$, 
Gerd P\"uhlhofer$^{4}$, Stefan Wagner$^{4}$ \\ 
for the H.E.S.S. Collaboration}
\shortauthors{H.E.S.S. Collaboration}
\afiliations{
$^1$School of Physics \& Astronomy, University of Leeds, Leeds LS2 9JT, UK\\
$^2$DAPNIA/DSM/CEA, CE Saclay, F-91191 Gif-sur-Yvette, Cedex, France\\
$^3$Max-Planck-Institut f\"ur Kernphysik, P.O. Box 103980, D 69029 Heidelberg, Germany\\
$^4$Landessternwarte, Universit\"at Heidelberg, K\"onigstuhl, D 69117 Heidelberg, Germany
}

\email{j.a.hinton@leeds.ac.uk}

\abstract{The rapidly varying non-thermal X-ray emission observed from Sgr A$^{\star}$ points to
particle acceleration taking place close to the supermassive black hole. The TeV
$\gamma$-ray source HESS\,J1745$-$290 is coincident with Sgr A$^{\star}$ and may be closely related
to the X-ray emission. Simultaneous X-ray and TeV observations are required to
elucidate the relationship between these two objects. Here we report on joint
H.E.S.S./Chandra observations in July 2005, during which an X-ray flare was detected.
Despite a factor $>10$ increase in the X-ray flux of Sgr~A$^{\star}$, no evidence
is found for an increase in the TeV $\gamma$-ray flux. We find that an increase
of the $\gamma$-ray flux of a factor 2 or greater can be excluded at a 
confidence level of 99\%. This finding disfavours scenarios in which the bulk
of the $\gamma$-ray emission observed is produced close to Sgr A$^{\star}$.
}

\begin{document}

\newcommand{\astar}{Sgr~A$^{\star}$}

\maketitle

\section{Introduction}

The existence of a supermassive ($3.6\pm0.3 \times
10^{6}$ solar mass) black hole at the centre of our galaxy has
been inferred using measurements of stellar orbits in the central parsec
(see e.g. ~\cite{GC:Eisenhauer05}). The supermassive black hole (SMBH)
is coincident with the faint radio source: Sgr~A$^{\star}$.  
The compact nature of Sgr~A$^{\star}$ has been demonstrated both by direct VLBI
measurements~\cite{GC:Shen05} and by the observation of X-ray and near
IR flares with timescales as short as a few minutes (see for
example~\cite{GC:Eckart06,GC:Porquet03}).  Variability on such short
timescales limits the emission region (via causality arguments) to within $<10$ Schwarzchild
radii of the black hole. X-ray flares from \astar\ have reached fluxes of $4\times10^{35}$ erg
s$^{-1}$, two orders of magnitude brighter than the quiescent
flux~\cite{GC:Porquet03, GC:Baganoff03}, and exhibit a range of
spectral shapes~\cite{GC:Porquet03}.  Several models exist for the origin of
this variable emission, all of which invoke non-thermal processes
close to the event horizon of the central black hole to produce a
population of relativistic particles.

Model independent evidence for the existence of ultra-relativistic 
particles close to Sgr~A$^{\star}$ can be provided by the observation
of TeV $\gamma$-rays from this source. Indeed, TeV 
$\gamma$-ray emission has been detected from the Sgr~A region 
by several ground-based instruments~\cite{GC:Whipple04,GC:CANGAROO,HESS:gc04,
GC:MAGIC06}. The most precise measurement of this source, HESS\,J1745$-$290,
are those made using the H.E.S.S. telescope array. The centroid
of the source is located $7'' \pm 14_{\mathrm{stat}}'' \pm 28_{\mathrm{sys}}''$ 
from Sgr~A$^{\star}$, and has an rms extension of  $<1.2'$~\cite{HESS:gcprl}.

TeV emission from \astar\ is expected in several models of 
particle acceleration in the environment of the black hole.
In some of these scenarios \cite{Levinson, AhNer1} TeV 
emission is produced in the immediate vicinity of the  
SMBH and variability is expected. In alternative scenarios
particles are accelerated at \astar\ but radiate in within
the central $\sim$10~parsec region~\cite{AhNer2}, 
or are accelerated at the termination shock of a wind 
driven by the SMBH \cite{AtDer}.
However, several additional candidate objects exist for the origin
of the observed $\gamma$-ray emission. The radio centroid of the 
supernova remnant (SNR) Sgr~A East lies $\sim 1'$ from
Sgr~A$^{\star}$, only marginally inconsistent with the position
of the TeV source given in \cite{HESS:gcprl}. Shell-type SNR are now well established TeV
$\gamma$-ray sources~\cite{HESS:rxj1713,HESS:velajnr}
and several authors have suggested Sgr A East as the origin of
the TeV emission (see for example \cite{Melia}). However,
recent improvements in the statistical and systematic 
uncertainties of the centroid of HESS\,J1745$-$290
effectively exclude Sgr~A East as the dominant 
$\gamma$-ray source in the region~\cite{VanEldik}.
The recently discovered pulsar wind nebula candidate G\,359.95-0.04~\cite{GC:Wang06}
lies only 9 arcseconds from \astar\ and can plausibly explain
the TeV emission~\cite{GC:Hinton07}. Particle
acceleration at stellar wind collision shocks within the central
young stellar cluster has also been hypothesised to explain
the $\gamma$-ray source~\cite{GC:Quataert05}. Finally, an origin of this source
in the annihilation of WIMPs in a central dark matter cusp has 
been extensively discussed~\cite{GC:Hooper04,GC:Profumo05,HESS:gcprl}.

Given the limited angular resolution of current VHE $\gamma$-ray 
telescopes, the most promising tool for identification of the
TeV source is the detection of \emph{correlated variability}
between the $\gamma$-ray and X-ray and/or NIR regimes.
A significant increase of the flux of HESS\,J1745$-$290 
simultaneous with a flare in wavebands with sufficient 
angular resolution to isolate \astar, would provide an 
unambiguous identification of the $\gamma$-ray source.
Therefore, whilst not all models for TeV emission from 
Sgr~A$^{\star}$ predict variability of the VHE source, 
coordinated IR/keV/TeV observations can be seen as a key
aspect of the ongoing program to understand the nature
of this enigmatic source.

\section{Observations \& Results}

A coordinated multi-wavelength campaign on \astar\ 
took place during July/August 2005. As part of this campaign 
observations with H.E.S.S. occurred
for 4-5 hours each night from the 27th of July to the 1st of August
(MJD  53578-53584). Four Chandra observations with IDs 5950-5954 took 
place between the 24th of July and the 2nd of August.
A search for flaring events in the X-ray data yielded
two significant events during the Chandra campaign, both during 
observation ID 5953 on the 30th of July. The second of these flares 
occurred during a period of H.E.S.S. coverage, at MJD 53581.94.

\begin{figure}[t]
\begin{center}
\noindent
\includegraphics [width=0.52\textwidth]{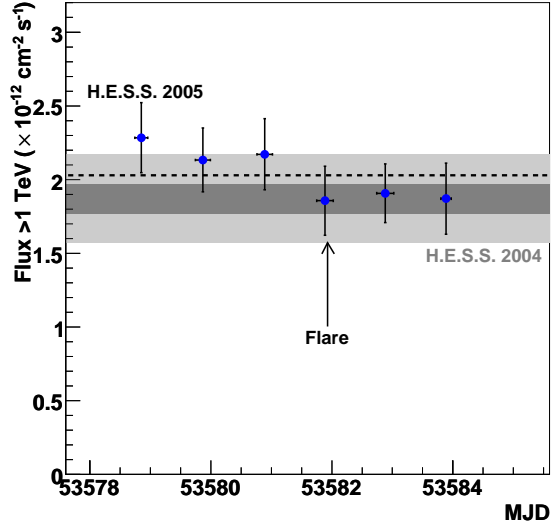}
\caption{Nightly $\gamma$-ray flux light-curve of HESS\,J1745$-$290 from the 
27th of July to the 1st of August 2005.
The H.E.S.S. data have typical thresholds of 150-300 GeV. 
The grey band shows the mean flux from 2004 observations as
published in \cite{HESS:gcprl}. Statistical (dark grey) and
systematic (light grey) errors are shown. The dashed line 
is a fit to the MJD 53578-53584 data.
}\label{fig0}
\end{center}
\end{figure}

The $\gamma$-ray data consist of 72 twenty-eight minute runs,
66 of which pass all the quality selection cuts described by 
\cite{HESS:crab}. All runs on the night of the X-ray flare pass
these cuts and in addition we find no evidence for cloud cover in the 
simultaneous sky temperature (radiometer) measurements 
(see \cite{HESS:crab,HESS:atmosphere}). 
These data were analysed using the H.E.S.S. standard 
\emph{Hillas parameter} based method 
described in \cite{HESS:crab}. An independent analysis based
on the \emph{Model Analysis} method described in \cite{deNaurois:Model} 
produced consistent results. Figure~\ref{fig0} 
shows a night-by-night TeV flux light-curve for this period.
There is no evidence for variations of the flux on day 
timescales and the mean $>1$ TeV $\gamma$-ray flux for this 
week of observations 
was $2.03 \pm 0.09_{\mathrm{stat}} \times 10^{-12}$ cm$^{-2}$ s$^{-1}$, 
consistent with the
average value for H.E.S.S. observations in 2004, 
$1.87 \pm 0.1_{\mathrm{stat}} \pm 0.3_{\mathrm{sys}} \times 10^{-12}$ 
cm$^{-2}$ s$^{-1}$~\cite{HESS:gcprl}.


\begin{figure*}
\begin{center}
\noindent
\includegraphics [width=1.0\textwidth]{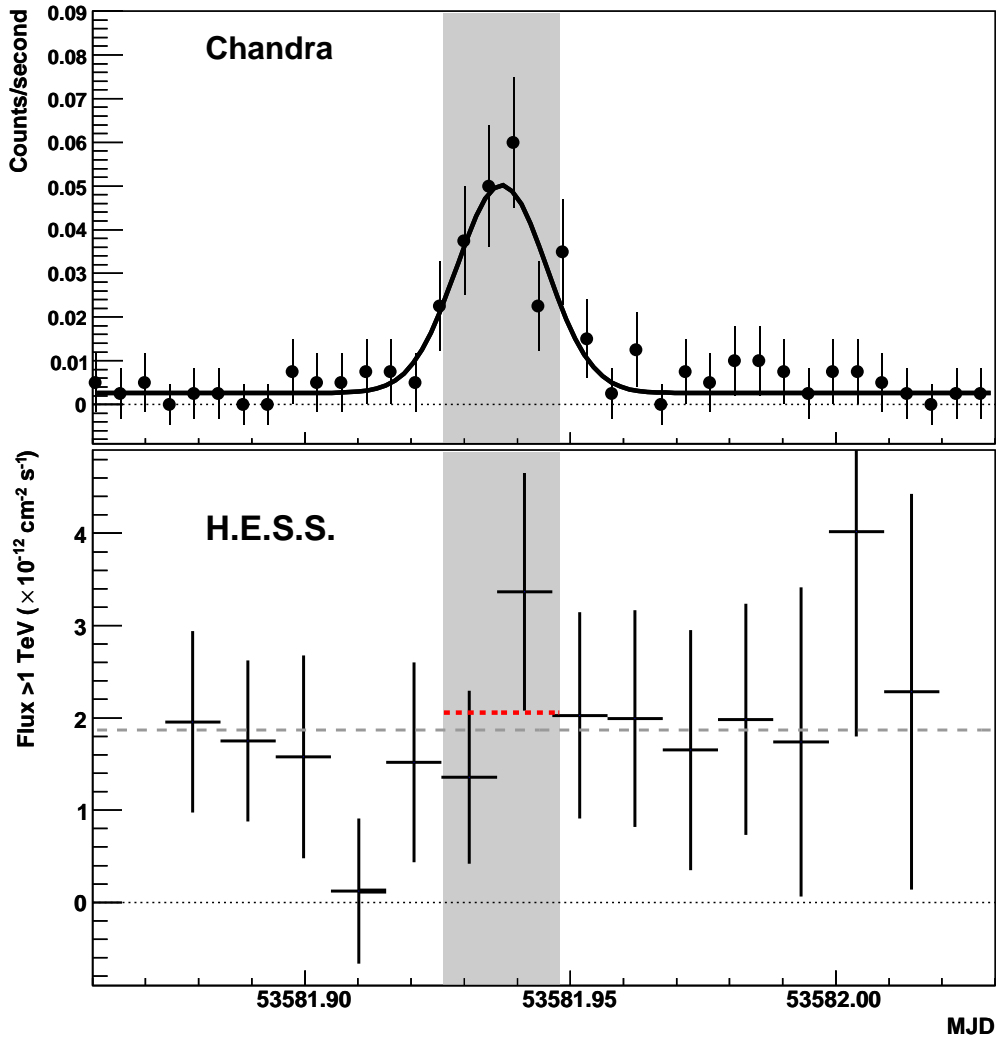}
\caption{X-ray and $\gamma$-ray light curves for the Galactic Centre on MJD 53581. 
  Top: Chandra 1-10 keV count rate in 400 second bins. The X-ray flare is well 
described by Gaussian (solid curve), the shaded region shows $\pm1.3\sigma$ of the
flare position.
Bottom: Very High Energy $\gamma$-ray light curve from H.E.S.S. in
15 minute bins. The long dashed line shows the historical flux
level \cite{HESS:gcprl}. The short dashed line indicates the mean 
TeV flux during the X-ray flare.
}	
\label{fig1}
\end{center}
\end{figure*}

Figure~\ref{fig1} shows the X-ray and $\gamma$-ray light curves 
for the night MJD 53581-2. There is a clear increase in the X-ray
flux of \astar\ with an excess of $103\pm10$ events with respect 
to the quiescent level. The time-profile of this excess is 
consistent with a Gaussian of rms $13.1\pm2.5$ minutes. The time
window for the $\gamma$-ray analysis is defined as the region within
$\pm1.3\sigma$ of the X-ray flare (containing 80\% of the signal).
The lower panel of Figure~\ref{fig1} shows the mean TeV flux
within this time window (grey shaded region) of 
$2.05\pm 0.76\,\times\,10^{-12}$ cm$^{-2}$ s$^{-1}$
as a short dashed line. This flux level is almost identical
to the mean flux level for this week of observations.
There is therefore no evidence for an increase in
$\gamma$-ray flux of HESS\,J1745-290 during the flare and a limit on the relative 
flux increase of $<$ a factor 2 is derived at the 99\% confidence 
level. In principle a (positive or negative) 
time lag might be expected between the X-ray and any associated 
$\gamma$-ray flare. The existence of a counterpart $\gamma$-flare with a 
flux increase $\gg2$ requires a lag of at least 100 minutes.

\section{Summary}

For the first time simultaneous TeV $\gamma$-ray observations 
have been presented for a period of X-ray activity of \astar.
The non-detection of an increase in the TeV flux provides an
important constraint on scenarios in which the source 
HESS\,J1745-290 is associated with the supermassive black hole.

\end{document}